\documentclass[usenatbib]{mn2e}
\usepackage{natbibmnfix, graphicx, times, amsmath, epsfig, amssymb}
\usepackage{amsmath}
\usepackage{verbatim}

\newcommand\lsim{\mathrel{\rlap{\lower4pt\hbox{\hskip1pt$\sim$}}
        \raise1pt\hbox{$<$}}}
\newcommand\gsim{\mathrel{\rlap{\lower4pt\hbox{\hskip1pt$\sim$}}
        \raise1pt\hbox{$>$}}}
\def\myputfigure#1#2#3#4#5%
{\vskip#5pt\makebox[0pt]{\hskip#2in
\includegraphics[width=#3\textwidth]{#1}}\vskip#4pt\hfill}

\begin{document}

\title[Focusing on Warm Dark Matter with Lensed High-redshift Galaxies]
      {Focusing on Warm Dark Matter with Lensed High-redshift Galaxies}
\author[F. Pacucci et al.]
{Fabio Pacucci$^1$ \thanks{fabio.pacucci@sns.it},
Andrei Mesinger$^1$, Zolt\'an Haiman$^2$ \\
$^1$Scuola Normale Superiore, Piazza dei Cavalieri, 7  56126 Pisa, Italy \\
$^2$Department of Astronomy, Columbia University, 550 West 120th Street, New York, NY, 10027, U.S.A.
}
             
\date{submitted to MNRAS}

\maketitle
             
\begin{abstract}
We propose a novel use of high-redshift galaxies, discovered in deep
Hubble Space Telescope (HST) fields around strong lensing
clusters. These fields probe small comoving volumes ($\sim10^3$ Mpc$^3$) at high magnification ($\mu\gsim 10$), and can detect
otherwise inaccessible ultra-faint galaxies. Even a few galaxies found
in such small volumes require a very high number density of collapsed
dark matter (DM) halos.  This implies significant primordial power on
small scales, allowing these observations to rule out popular
alternatives to standard cold dark matter (CDM) models, such as warm
dark matter (WDM).  In this work, we analytically compute WDM halo
mass functions at $z=10$, including the effects of both particle
free-streaming and residual velocity dispersion.  We show that the two
$z\approx10$ galaxies already detected by the Cluster Lensing And Supernova survey with Hubble (CLASH) survey are sufficient to
constrain the WDM particle mass to $m_x >$ 1 (0.9) keV at 68\% (95\%) confidence limit (for a thermal relic relativistic at decoupling).
This limit depends only on the WDM halo mass function and, unlike
previous constraints on $m_x$, is independent of any
astrophysical modeling.  The forthcoming HST Frontier Fields can
significantly tighten these constraints.
\end{abstract}


\setcounter{footnote}{1}

\section{Introduction}
\label{sec:intro}
The $\Lambda$CDM cosmological model, in which structure formation proceeds in a hierarchical manner driven by pressure-less cold dark matter (CDM), has been remarkably successful in predicting the matter distribution on large scales.  It accurately describes the statistical properties of the cosmic microwave background (CMB), cluster abundances, galaxy clustering and the cosmic web (e.g. \citealt{Tegmark2006, Benson2010, Komatsu2011, Hinshaw2012}).

However, observations of low-redshift galaxies over the past decade have suggested that CDM predicts too much power on small scales. For instance, CDM $N$-body simulations contain  more satellite 
galaxies than are observed both around our galaxy (the so-called ``missing satellite problem''), in voids, and in the field (e.g. \citealt{Moore1999, Klypin1999, Diemand2008, Springel2008, Papastergis11, Ferrero2012}).
The inner profiles of individual dwarf galaxies are too shallow, compared with CDM predictions (e.g. \citealt{Moore1994, deBlok2001, Maccio2012, Governato2012}).
Furthermore, CDM simulations result in a population of massive, concentrated Galactic sub-halos that are inconsistent with kinematic observations of the bright Milky Way satellites \citep{Boylan2012}.

An apparent solution can be found by appealing to the baryons.  Baryonic feedback caused by supernovae explosions and heating due to the UV background may suppress the baryonic content of low-mass halos and make their inner density profile shallower (e.g. \citealt{Governato2007, Mashchenko2008, Busha10, Sobacchi2013, Teyssier2013}).  Nevertheless, it is not clear if baryonic feedback provides a satisfactory match to all observations, even when its implementation is ``tuned'' relatively arbitrarily (e.g. \citealt{Boylan2012, Garrison2013, Teyssier2013}).

An alternative explanation might be found if dark matter consisted of lower mass\footnote{Throughout this work, we report warm dark matter particle masses, $m_x$, assuming thermal relics which were relativistic at decoupling (e.g. neutralino, gravitino).  Results can be translated for non-thermal relics, such as the sterile neutrino, by equating their free-streaming lengths, though care should be taken in interpreting the effects of residual velocities in those cases (see below).} ($\sim$ keV) particles, so-called warm dark matter (WDM; \citealt{Blumenthal1982}).
Given their lower mass with respect to cold dark matter, WDM particles remain relativistic for a longer time and they are able to free-stream out of potential wells, smearing out small-scale primordial perturbations. In addition, their velocity dispersion acts as an effective pressure which suppresses growth of perturbations on scales smaller than a characteristic Jeans mass.  Therefore, small-scale structure is dramatically suppressed in WDM models, and models with WDM masses of $m_x \sim$ keV might provide a better match to observations of local galaxies (e.g. \citealt{BOT01, Khlopov2008, Valageas2012, Kang2013, Angulo2013, Viel2013}; though see \citealt{Maccio12_catch22} and \citealt{Shao2013} who argue that the required WDM masses are low enough to be ruled out by current observations).

The most powerful test-bed for these scenarios is the high-redshift Universe. Structure formation in WDM models 
(or in any cosmological model with an equivalent power-spectrum cut-off) is exponentially suppressed on 
small scales. Since in a CDM dominated Universe structures form hierarchically, these small halos are expected to host the first galaxies\footnote{Once larger scales go nonlinear at lower redshifts, in principle,
low-mass halos can form in WDM models via top-down fragmentation,
spoiling such direct inferences (e.g. \citealt{BOT01}; however, such
fragmentation is exceedingly difficult to simulate reliably, see \citealt{WW2007}).  At sufficiently high redshifts, before {\em
any} scales go nonlinear, no halos of any mass exist.}. 
Therefore, the mere presence of a galaxy at high-redshift ($z\gsim 10$) can set strong lower limits on the WDM particle mass (e.g. \citealt{BHO01,MPH05}).

Various high-$z$ observations have already been used to constrain WDM properties.
WDM masses lower than $m_x \approx$ 3 keV would smear out the Lyman alpha forest seen in high-$z$ quasar spectra (\citealt{Viel2008}; Viel et al., in preparation).  Somewhat weaker limits, $m_x \gsim 0.75$ keV are obtained by simultaneously reproducing the stellar mass function and the Tully-Fisher relation \citep{Kang2013}.  Requiring that reionization is completed by $z\gsim6$ yields a similar limit, as does the need to grow $\sim 10^9~{\rm M_\odot}$ supermassive black holes by this redshift \citep{BHO01}. 
However, most of these limits are strongly affected by a degeneracy between astrophysical processes (involving baryons) and the dark matter mass.  Recently, \cite{DeSouza2013} set robust limits of $m_x > 1.6-1.8 $ keV ($95\%$ confidence level) from the number of high-$z$ gamma ray bursts (GRBs) observed with {\it Swift}, using a Bayesian likelihood analysis as well as conservative assumptions to minimize degeneracy with the astrophysics.

{\it In this work we show how the abundances of lensed, very high-redshift galaxies can be used to robustly probe primordial small-scale power, as typified by WDM models.}  Lensing facilitates the detection of ultra-faint high-$z$ galaxies in fairly small volumes. The mere detection of a single halo in this volume would imply a very high number density, setting a tight upper limit on the WDM particle mass, {\it completely independent of astrophysical degeneracies}.
Currently, there are two highly magnified galaxy candidates at $z\sim10$, discovered as part of the Cluster Lensing And Supernova survey with Hubble (CLASH; \citealt{Postman2012, Zheng12, Coe2013}).
MACS1149-JD is at $z = 9.6 \pm 0.2$, magnified by a factor of $\mu \sim 15$. 
MACS0647-JD at $z = 10.7_{-0.4}^{+0.6}$ has three images, one of which is magnified by a factor of $\mu \sim 8$.

The outline of this paper is as follows. In $\S 2$ we discuss how we derive dark matter halo mass functions in both CDM and WDM models.
In $\S 3$ we present the resulting mass functions at $z=10$, comparing them with recent CLASH observations.
In $\S 4$ we discuss future improvements, while in $\S 5$ we report our conclusions.
Throughout, we adopt recent Planck cosmological parameters: $(\Omega_m, \Omega_{\Lambda}, \Omega_b, h, n_s, \sigma_8 )= (0.32, 0.68, 0.05, 0.67, 0.96, 0.83)$, and assume WDM is a fermionic spin $1/2$ particle.  All scales are quoted in comoving units, unless otherwise indicated.

\section{Mass functions in CDM and WDM}
\label{sec:results}

In WDM models the growth of density perturbations is suppressed on small scales. This is due to a combination of two different effects: (i) particle free-streaming during the radiation-dominated era smears-out small-scale structure, altering the effective transfer function of the matter power spectrum (e.g. \citealt{BOT01}); and (ii) the stochastic residual velocity dispersion of particles, although redshifting away as $\propto(1+z)$, suppresses the initial growth of DM perturbations, acting as an effective pressure term.  Below, we describe how we account for these two effects in constructing analytic mass functions, closely following the procedure recently outlined in \cite{Benson2013}.

\subsection{Power spectrum cut-off from WDM free-streaming}
\label{sec:ps}

The free-streaming scale for a thermal relic, defined as the scale at which the linear perturbation amplitude is suppressed by a factor of two, is given by \citep{BOT01}
\begin{equation}
R_{\rm fs}\approx 0.31 \left( \frac{\Omega_x}{0.3} \right)^{0.15} \left( \frac{h}{0.65} \right)^{1.3} \left( \frac{\rm keV}{m_x} \right)^{1.15} \frac{\rm Mpc}{h}~,
\end{equation}
where $\Omega_x$ is the fraction of the critical energy density contained in WDM, and $h$ is the Hubble constant in units of 100 km s$^{-1}$ Mpc$^{-1}$. The free-streaming scale corresponding to $m_x=1$ keV is $R_{\rm fs}\approx 0.52$ Mpc. The  corresponding modification of the matter power spectrum can be computed by multiplying the CDM power spectrum $P_{\rm CDM}(k)$ by an additional scale-dependent transfer function \citep{BOT01}:
\begin{equation}
P_{\rm WDM}(k) = P_{\rm CDM}(k) \left[1 + (\epsilon k)^{2\mu}\right]^{-5\mu} ~,
\end{equation}
where $\mu=1.12$ and 
\begin{equation}
\epsilon=0.049\left(\frac{\Omega_x}{0.25}\right)^{0.11}\left(\frac{m_x}{\rm keV}\right)^{-1.11} \left(\frac{h}{0.7}\right)^{1.22}\rm{h^{-1}Mpc} \,\, .
\end{equation}
The root variance in the $z=0$ linearly-interpolated matter field on scale $M$ is then computed according to:
\begin{equation}
\sigma^2(M) = \frac{1}{2\pi^2}\int_{0}^{\infty} 4\pi k^2 P(k)W^2(k|M)\, dk
\end{equation}
We use a sharp k-space filter window function in the case of WDM,
\begin{equation}
 W(k|M) =
  \begin{cases}
   1 & \text{if } k \leq k_s(M) \\
   0 & \text{if } k > k_s(M) 
  \end{cases}
\end{equation}
\noindent relating the cut-off mode $k_s$ to the spatial scale $R=[3 M / ( 4 \pi \bar{\rho})]^{1/3}$ through $k_s=2.5/R$.  \cite{Benson2013} show that this choice of window function and normalization accurately reproduces WDM mass functions from $N$-body simulations.

\subsection{Effective pressure from WDM residual velocities}
\label{sec:pressure}

Structure formation in WDM models will be further suppressed by the residual velocity dispersion of the WDM particles,
which delays the growth of perturbations. 
This effect can be incorporated in the halo mass function by raising the critical overdensity threshold required for collapse, $\delta_c(M, z)$. 
Using spherically symmetric hydrodynamics simulations, and exploiting the analogy between the WDM effective pressure and gas pressure, \cite{BHO01} computed the modified $\delta_c(M,z)$ to be used in the excursion-set random walk procedure.  They show that this effective pressure can be of comparable importance to the power spectrum cut-off above, in suppressing small-scale structures. \cite{Benson2013} showed that the results of \cite{BHO01} can be well fitted by the following functional form:
\begin{align}
\delta_{\rm c,WDM}(M,z) =& \,\delta_{\rm c,CDM}(t)\Big\{h(x)\frac{0.04}{\exp(2.3x)} + \\
&+[1-h(x)] \exp\Big[\frac{0.31687}{\exp(0.809x)}\Big]\Big\}.
\end{align}
Here $x=\log(M/M_J)$ and $M_J$ is an effective WDM Jeans mass, i.e., the mass scale below which collapse is significantly delayed by the pressure:
\begin{align}
M_{\rm WDM}\sim&\,3.06 \times 10^{8} \Big(\frac{\Omega_x h^2}{0.15}\Big)^{1/2} \Big(\frac{m_x}{\rm{keV}}\Big)^{-4}\times \\ 
&\times\Big(\frac{1+z_{\rm{eq}}}{3000}\Big)^{1.5}\Big(\frac{g_x}{1.5}\Big)^{-1} \rm{M_{\odot}},
\end{align}
where $z_{eq}$ is the redshift of radiation-matter equality, $g_x$ is the effective number of degrees of freedom of WDM, and
\begin{equation}
h(x)=\left\{1+\exp\left[\frac{x+2.4}{0.1}\right]\right\}^{-1} ~ .
\end{equation}
\cite{Benson2013} also note that $\delta_{\rm c,WDM}(M,z)$ should be multiplied by an additional factor of 1.197, in order to compensate for the non-standard normalization of the window function above.  This normalization is motivated by requiring  WDM and CDM mass functions to converge at high masses. In order to compute the WDM mass function, one can record the first crossing probability, $f(\sigma^2)$, averaging over many random walks (\citealt{BHO01, MPH05}).  Instead we use the faster recursive procedure described in Appendix A of \cite{Benson2013}, to which we refer the reader for further details.

\subsection{CDM mass functions}

We adopt a standard analytic formula, which is known to accurately fit $N$-body results  (\citealt{ST99, Jenkins01}), to
compute CDM mass functions:
\begin{equation}
f_{\rm ST}=A\sqrt{\frac{2a_1}{\pi}}\times \left[1+\left(\frac{\sigma^2}{a_1\delta_c^2}\right)^p\right]\times\frac{\delta_c}{\sigma}\exp{\left(-\frac{a_1\delta_c^2}{2\sigma^2}\right)}
\end{equation}
where $f_{\rm ST}$ is the fraction of mass contained in halos with masses greater than $M$, and $A=0.322$, $a_1=0.707$, $p=0.3$, $\delta_{\rm c}=1.686$.
The mass fractions, $f_{ST}$, can be related to the halo number density $n_{\rm ST}(M,z)$ through
\begin{equation}
f_{\rm ST}=\frac{M}{\rho_m}\frac{dn_{\rm ST}(M,z)}{d\ln{\sigma^{-1}}}
\end{equation}
where $\rho_m$ is the total mass density of the background Universe.

\section{Results}
\label{sec:res}

\begin{figure}
\vspace{-1\baselineskip}
\hspace{-0.5cm}
\includegraphics[width=0.55\textwidth]{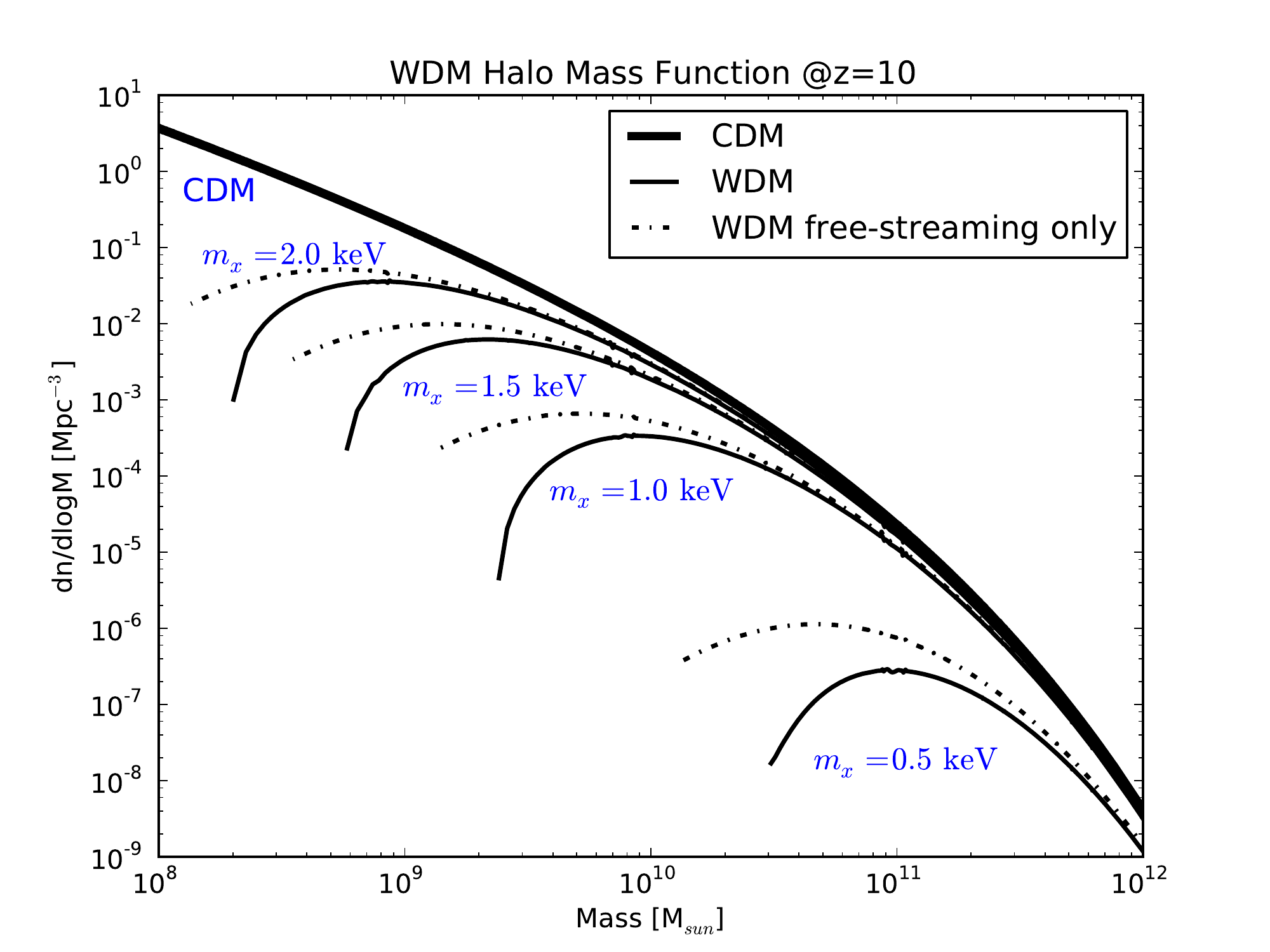}
\caption{Halo mass functions in CDM (thick solid line) and WDM (solid lines) at $z=10$. The dot-dashed lines correspond to WDM models where only the free-streaming effect has been accounted for (i.e. the effective pressure from the residual velocity dispersion of the WDM particles is neglected).}
\label{fig:mass_functions}
\vspace{-1\baselineskip}
\end{figure}

The resulting CDM and WDM halo mass functions at $z=10$ are shown in Figure \ref{fig:mass_functions}, for four different values of the WDM particle mass: $m_x$ = 0.5, 1.0, 1.5, 2.0 keV. Solid lines include both of the effects mentioned above (free-streaming and effective pressure), while dashed lines only include the free-streaming. The latter can be considered as strict, highly-conservative, particle-type independent estimates of the impact of WDM on high-$z$ mass functions\footnote{The effects of free-streaming (\S \ref{sec:ps}) are straightforward to compute analytically and to include in $N$-body simulations as a modification of the initial power spectrum (e.g. \citealt{BOT01}).  Analytic WDM halo mass functions are in agreement with $N$-body codes, when compared just including free-streaming and accounting for artificial fragmentation of $N$-body filaments \citep{WW2007}.  Furthermore, the resulting modification of the power spectrum can be computed for any given WDM particle, and so results for our fiducial thermal relic mass, $m_x$,  can be translated to masses of other WDM particles (e.g. sterile neutrinos), by equating their corresponding streaming lengths.  On the other hand, the effects of WDM velocity dispersion (\S \ref{sec:pressure}) are more uncertain, as they require accounting for intra-particle dispersion in $N$-body simulations.  Thus our prescription for including the effects of residual velocities on structure formation is not as well tested as the effects of free-streaming.
  Furthermore, our method is motivated by the results of \citet{BHO01}, who made the analogy of WDM velocity dispersion to gas temperature in their simulations.  This analogy is only well defined in the case of a thermal relic WDM particle.}.
The thick solid line is the standard CDM halo mass function. The exponential suppression of low-mass halos in WDM models is evident in the figure, with the abundance of $\sim10^8 \rm{M_\odot}$ halos dramatically reduced for $m_x \lsim 2$ keV.  The effective pressure from WDM velocity dispersion makes this mass function truncation much more dramatic (c.f. \citealt{BHO01, MPH05}).  Halos  with $M\approx10^8 \rm{M_\odot}$ correspond to the lowest masses in which cooling by atomic hydrogen is efficient, suggesting that abundances of $z\sim10$ star-forming galaxies are sensitive to $m_x \lsim 2$--3 keV (e.g., \citealt{DeSouza2013}).

\begin{figure}
\vspace{-1\baselineskip}
\hspace{-0.5cm}
\includegraphics[width=0.55\textwidth]{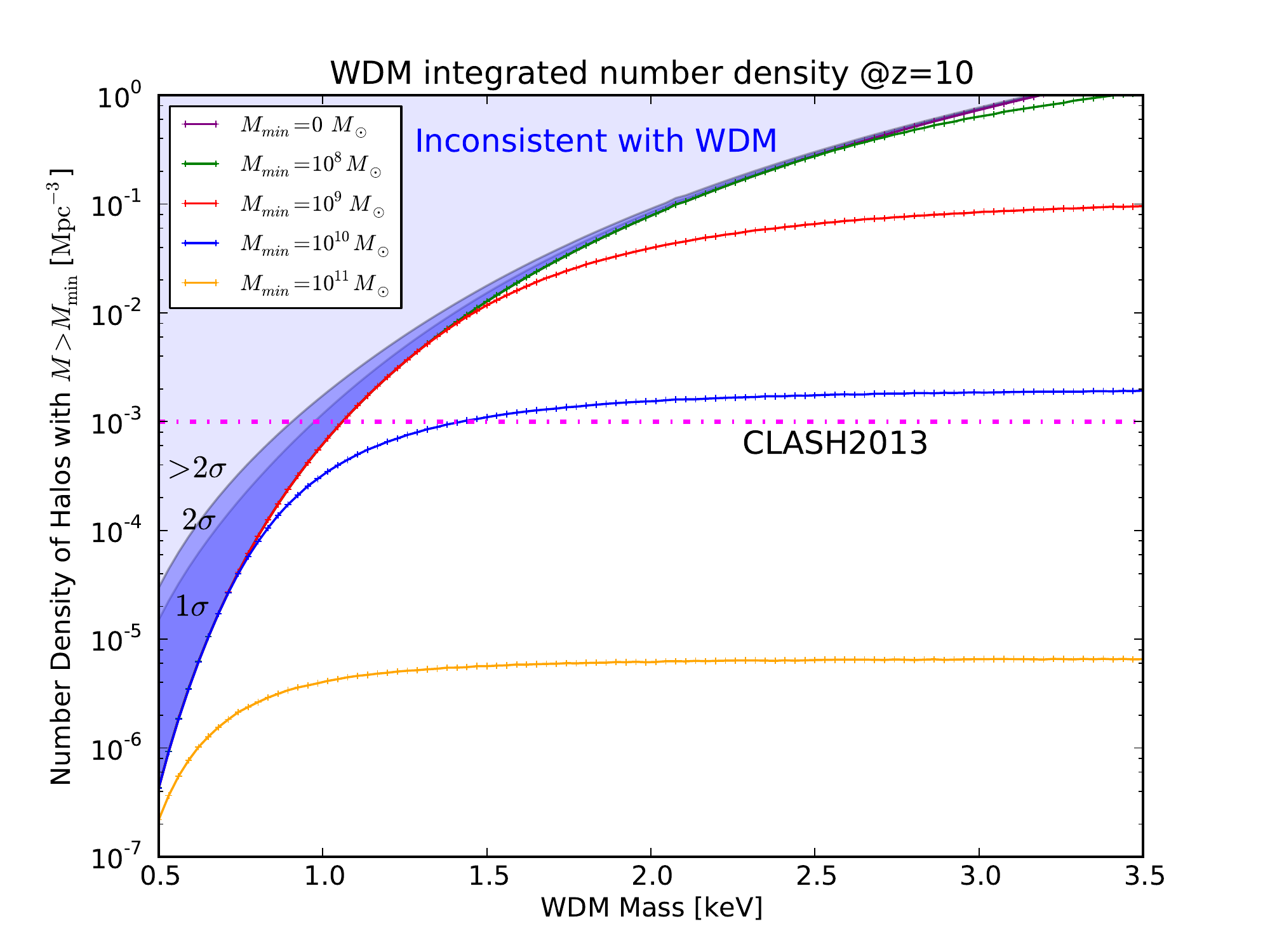}
\caption{Integrated number densities of WDM halos as a function of the particle mass, $m_x$.  Abundances were computed down to a minimum halo mass, $M_{\rm min}$.  The shaded region corresponds to abundances higher than those integrating all the way down to $M_{\rm min}=0 \, \rm{M_\odot}$ (i.e. with no lower limit on the halo mass), and are strictly ruled out.
The number density implied by the two lensed CLASH candidates is demarcated with a horizontal purple line.
The darker shaded regions corresponds to the 1 and 2 sigma Poisson uncertainties on the number predicted by the $M_{\rm min}=0\, \rm{M_{\odot}}$ models for the CLASH volume cited in the text. The striking difference between the 1 sigma and 2 sigma levels is due to the log-scale used.}
\label{fig:num_den}
\vspace{-1\baselineskip}
\end{figure}

By integrating the WDM mass functions in Figure  \ref{fig:mass_functions} above a low-mass limit, $M_{\rm min}$, one obtains the total number density of halos at $z=10$ with mass $M > M_{\rm min}$, shown in Figure \ref{fig:num_den}.  This figure is the main result of this work.  As an example of its use, let us consider halo abundances in WDM models with $m_x < 3$ keV. We note from the figure that halos, {\it of any mass}, cannot be more abundant than $\approx0.8$ Mpc$^{-3}$ if $m_x\lsim 3$ keV.  This ``forbidden region'' is highlighted in blue, and {\it is completely independent of the physics of the baryons}.   
In principle, stronger constraints could be obtained, at the expense of astrophysical modeling uncertainties, if such modeling requires host halos of detected galaxies to be higher than $M_{\rm min}\gsim 10^{9-10} {\rm M_\odot}$. For example, considering that star-forming galaxies at this redshift\footnote{The very first galaxies are expected to be hosted by less massive halos, in which gas cooled via the molecular cooling channel.  However, the H$_2$-disassociating background likely sterilized star formation in these galaxies long before $z=10$ (e.g., \citealt{HL97, MO12}).} are unlikely to be hosted by halos less massive than the corresponding atomic cooling threshold (i.e. with virial temperatures less than $\approx 10^4$ K, corresponding to $M_{\rm min} \approx 10^8 \rm{M_\odot}$ at $z=10$), one can set only a modestly stronger\footnote{Since our WDM mass functions truncate abruptly towards low masses, there are virtually no halos with masses lower than $\lsim 10^8 \rm{M_\odot}$, even for particle masses as high as $m_x\sim3$ keV.} upper limit on the {\it abundance of star-forming galaxies} corresponding to the green curve, $\lsim0.6$ Mpc$^{-3}$ for $m_x\lsim 3$ keV.  Further including some modeling of the halo mass $\rightarrow$ galaxy luminosity, combined with the telescope sensitivity limits, allows one to obtain estimates of the {\it abundances of observable galaxies}\footnote{Note that the observed Lyman break galaxy candidates at $z\sim6$--8 are expected to be hosted by $M_{\rm min}\gsim10^{10}-10^{11} \rm{M_\odot}$ halos, when abundances are computed within a CDM framework, and these relatively large halo masses are generally consistent with the observed Lyman $\alpha$ fluxes (e.g. \citealt{Bouwens08, Labbe10}).}.

We now consider the current limits available from the two CLASH lensing candidates.  The total effective volume, $V_{\rm eff}(\mu)$, is a function of the magnification factor, $\mu$, with higher magnification factors corresponding to smaller effective volumes. Currently the first 12 clusters (out of a total of 25) have been processed, with $V_{\rm eff}(\mu)$ computed. 
The two lensing candidates from CLASH, along with the 12-cluster comoving volume at that magnification,  $V_{\rm eff}(\mu)$, are listed in the table below  (Moustakas, L. A., private communication):
\begin{center}
    \begin{tabular}{ | c | c | c |}
    \hline
     Object ID & $\mu$ & $V_{\rm eff}(\mu) [{\rm Mpc}^3]$ \\ \hline
    MACS1149-JD &15 & $\sim700$ \\ \hline
    MACS0647-JD & 8 & $\sim2000$ \\ \hline
    \end{tabular}
\end{center}
The effective volumes are very small, and hence the implied abundances relatively large.  If we conservatively take the volume corresponding to the lower magnification, the two candidates give a number density of $2/2000$ Mpc$^{-3}$, or:
\begin{equation}
n_{\rm tot} \sim 10^{-3} ~ {\rm Mpc}^{-3} ~.
\end{equation}
This total number density is shown by the dashed horizontal purple line in Figure $\ref{fig:num_den}$.

The intersection of this horizontal line with the edge of the ``forbidden region'' provides the robust constraint of $m_x > 1$ keV.  For this constraint, we have included the 1-$\sigma$ Poisson errors (68\% confidence limit; C.L.) on the expectation value for the number of predicted halos in the total CLASH volume in the $M_{\rm min}=0$ models (shown as the dark blue shaded region in Fig. \ref{fig:num_den}).  The corresponding 2-$\sigma$ constraint (95\% C.L.) is $m_x > 0.9$ keV (medium blue shaded region).
 This constraint is competitive with other current WDM constraints discussed above.  More importantly however, {\it it is the only constraint on the WDM particle mass completely independent of astrophysics}.

For reference, we have also computed the WDM halo number densities ignoring the effects of peculiar velocities.  We find that ignoring the effects of peculiar velocities (i.e. only accounting for free-streaming) mildly degrades the above constraints from CLASH, to $m_x > 0.9$ keV (68\% C.L.).

\begin{figure}
\vspace{-1\baselineskip}
\hspace{-0.5cm}
\includegraphics[width=0.55\textwidth]{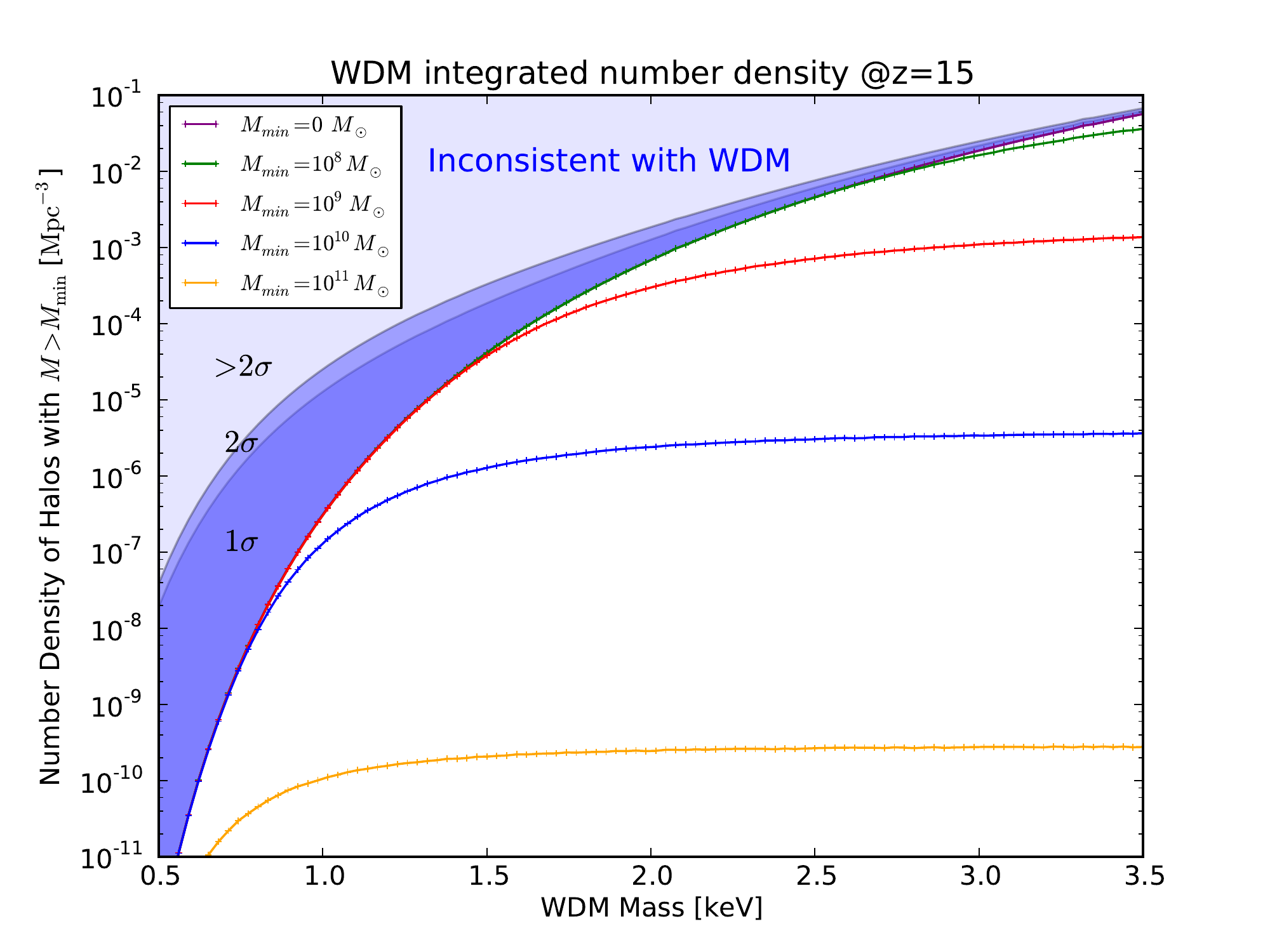}
\caption{Same as Fig. \ref{fig:num_den}, but computed at $z=15$.}
\label{fig:num_den_15}
\vspace{-1\baselineskip}
\end{figure}

Estimates of the halo number densities at higher redshifts could provide even tighter constraints on the WDM particle mass. For example, if an object at $z\sim15$ was discovered in a similar volume, the WDM constraint would strengthen to $m_x>2.2 \, \rm{keV}$ (see Fig. \ref{fig:num_den_15}).  Or looking at it from a different perspective, for a $m_x=1$ keV model, one would need to cover a comoving volume which is 10$^5$ times larger than that of CLASH, in order to find a single object at $z\sim15$.

We note also that MACS0647-JD has multiple images, and can therefore confidently be considered to be a real $z\sim10$ galaxy.  Using the number density implied by just this one galaxy, $n_{\rm tot} \sim 5\times10^{-4} {\rm Mpc}^{-3}$, our constraints would degrade to $m_x > 0.8$ keV.

\section{Future potential}

Our framework can be used with future datasets. Presently the CLASH survey has full data on 23 out of a total 25 lensing clusters, with analysis completed on 12 out of the full 25 (Moustakas, L. A., private communication).  Our results can be updated when the final CLASH number density becomes known.  Future, deeper cluster surveys can also improve on the observed number density by detecting even fainter galaxies, closer to the WDM cut-off.  
In particular, the HST Frontier Fields\footnote{See http://www.stsci.edu/hst/campaigns/frontier-fields} will target 4-6 lensing clusters, and obtain images $\approx 3$ magnitudes deeper than CLASH, probing an order of magnitude fainter galaxies, residing in lower-mass halos.

For a more robust constraint, the full lensing geometry of each cluster can be modeled, accounting for both the error bars on the modeled $V_{\rm eff}(\mu)$, the redshift evolution of the mass function across this volume (since the lensed volume is not a simple comoving cube around $z=10$), as well as the correlation function of the halos (as the sampling statistics are not quite Poisson, due to the clustering of halos). 
We defer these to future work.

Tighter constraints on $m_x$ can also be obtained at the cost of including astrophysical uncertainties.  Our constraints above are based on the presence of {\it any halos} (i.e. using $M_{\rm min}=0 \,\rm{M_\odot}$).  As mentioned above, if we consider that luminous galaxies must be hosted by relatively large halos, our constraints can improve.
   Relating the galaxy luminosity to the host halo mass is very uncertain, though one can make conservative choices for $M_{\rm min}$.  
This is because, as shown in Fig.~\ref{fig:mass_functions}, the peak halo mass in WDM models can be significantly below $10^{10}~{\rm M_\odot}$ - if sufficiently luminous objects are found, their corresponding host halo masses implied by abundance matching would require uncomfortably low mass-to-light ratios in WDM models.  In the extreme limit, the mass-to-light ratio could be lower than obtained by even allowing all of the baryons to form stars in a single burst.
   Likewise, if the observed color of a lensed galaxies implies an age older than $\gsim 100$ Myr, this would imply the existence of DM halos at $z\gsim 12$, again strengthening constraints.
Nevertheless, pushing limits beyond $m_x \gsim$ 2--3 keV would require going to even higher redshifts, since at those particle masses even the $M_{\rm min}\approx10^8$--$10^9 \rm{M_\odot}$ abundances asymptote to CDM values (see also Fig. 1 in \citealt{DeSouza2013}).

\section{Conclusions}
\label{sec:conclusion}

Structure formation in WDM models (or any similar model with a steep
power-spectrum cut-off) is dramatically suppressed at small-scales.
Low-mass halos can only form via top-down fragmentation at late times.
This leads to a dramatic difference between halo abundances in CDM and WDM models at the high redshifts, with the Universe becoming increasingly empty as the WDM particle mass, $m_x$, is decreased.

 In this work, we illustrate how the high implied abundances of lensed galaxies at $z\sim10$ can be used to set robust constraints on $m_x$.  Using two $z\sim10$ galaxies observed to date by CLASH, we set lower limits of $m_x > 1$ (0.9) keV at 68\% (95\%) C.L..  This limit is the first constraint on $m_x$ strictly independent of any astrophysical degeneracies
 -- the only modeling required is that of the DM halo mass function (which should be verified in future numerical WDM simulations, targeting $z\gsim10$) and of the effective volume of the lensing observations.

\vspace{+1cm}
We thank Leonidas A. Moustakas and other members of the CLASH survey team for providing current estimates of the effective lensed survey volume.  This was was supported by the NSF under grant AST-1210877.


\bibliographystyle{mn2e}
\bibliography{ms}

\end{document}